\documentclass{kapproc} 
\usepackage{graphics}
\usepackage{epsfig}

\kluwerbib

\begin{document}

\articletitle{Magnetic Fields in the envelopes of late-type stars}

\articlesubtitle{Circular Polarization of H$_2$O Masers}

\author{W.H.T.~Vlemmings, P.J.~Diamond, H.J.~van~Langevelde}
\affil{Sterrewacht Leiden\\
Niels Bohr weg 2, 2300 RA Leiden, The Netherlands}
\email{vlemming@strw.leidenuniv.nl}

\begin{abstract}
We present the first circular polarization measurements of
circumstellar H$_2$O masers around a sample of late-type stars. These
observations are used to obtain the magnetic field strength in the
H$_2$O maser region with both an LTE and non-LTE analysis. We find
fields from a few hundred milliGauss up to a few Gauss, indicating a
solar-type $r^{-2}$ dependence of the magnetic field on the distance.
No linear polarization is detected to less than 1\%.
\end{abstract}

\begin{keywords}
masers - polarization - stars: circumstellar matter - stars: magnetic
fields - techniques: interferometric
\end{keywords}

\section*{Introduction}

The role of magnetic fields in the late stages of stellar evolution is
still unclear. Blackman et al. (2001) have shown that AGB stars could
produce fields of several hundreds of Gauss. Such strong fields can
play an important role in driving stellar winds and shaping the
outflows.

 Until recently, information on the magnetic field in the
circumstellar envelopes was obtained by polarimetric observations of
SiO masers at $\approx~2-4~R_*$ from the central star, and OH
maser at a distance of $1000 - 10000$~AU. The SiO observations
indicated fields of 5-10~Gauss for Mira stars and up to 100~Gauss for
supergiants (e.g. Barvainis et al., 1987). However, using a non-Zeeman
interpretation of the observed circular polarization, the magnetic
fields could be a factor 1000 less (Wiebe \& Watson, 1998). OH maser
observations indicated field strengths of 1-2~mG (e.g. Szymczak \&
Cohen, 1997).

 Now we have been able to determine the magnetic field strengths in
the intermediate region, at a few hundred AU, where the H$_2$O masers
occur. Although H$_2$O is a non-paramagnetic molecule, it has been
possible to observe the circular polarization on some of the strongest
circumstellar H$_2$O maser features. We have used both the LTE
analysis presented in Vlemmings et al. (2001) and non-LTE
models based on the models presented in Neduloha \& Watson
(1990). A full description of the analysis methods, observations and
results are presented in Vlemmings et al. (2002).

\section{Observations}

We have observed 4 late-type stars with the VLBA, the supergiants
S~Per, VY~CMa and NML~Cyg, and the Mira variable star U~Her. To get
the highest spectral resolution, required for the circular
polarization measurements, the data were correlated twice. Once with
modest spectral resolution ($0.1$ km/s), to get all 4 polarization
combinations (RR, LL, RL and LR), and once with high resolution
($0.027$ km/s), with only RR and LL. The calibration was mainly
performed on the modest spectral resolution data and the solutions
were copied and applied to the high resolution data. This data set was
then used to produce circular polarization and total intensity image
cubes.

\begin{figure}
\resizebox{\hsize}{!}{\includegraphics{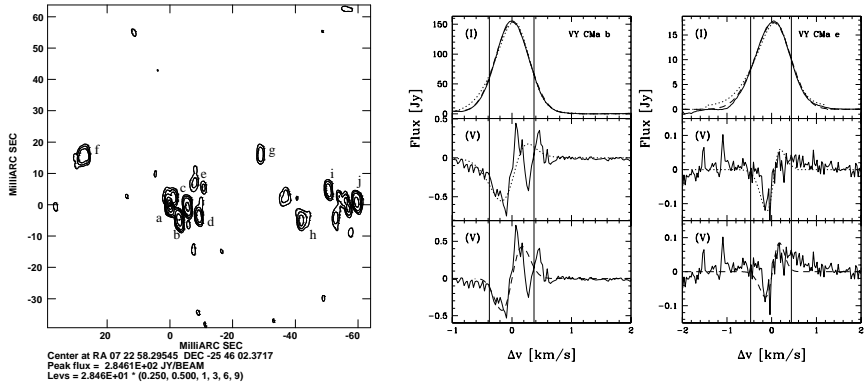}}
\caption{(left) Total intensity image of the H$_2$O maser features
  around VY~CMa. (right) Total power (I) and V-spectra for selected
  maser features of VY~CMa. The bottom panel shows the best fitting
  LTE model (dashed), the middle panel shows the best fitting non-LTE
  model (dotted).}
\end{figure}

\section{Results}

 We have examined the strongest H$_2$O maser features around the 4
stars observed. Circular polarization down to $0.2\%$ of the total
intensity was detected on $\approx 50\%$ of the brightest maser
features. An example of the features around VY~CMa is shown in Fig.1.
We rule out any systematic effects as a cause of the observed
spectrum, because various percentages of circular polarization are
observed as well as different directions.  No linear polarization was
detected above the limit of $\approx 0.5\%$.

 The magnetic field strengths were determined with both the LTE and
the non-LTE method. The LTE method predicts the circular polarization
spectrum to be directly proportional to the derivative of the total
power spectrum. We found that the observed spectra were narrower,
which can only be explained with the non-LTE analysis. The non-LTE
field strengths are $\approx 40\%$ lower than those determined by the
LTE method.

From the observations and analysis, we estimate the magnetic field
strengths in the H$_2$O maser region to be $\approx 200$~mG for S~Per
and VY~CMa. The field around NML~Cyg is $\approx 500$~mG, while the
Mira variable U~Her shows a much higher field of $\approx 1.5$~G.

\section{Conclusions}

 Our results favor the non-LTE approximation and because we do not
detect any linear polarization, a non-Zeeman interpretation is also
highly unlikely. The lack of linear polarization can be easily
explained in the non-LTE case, because linear polarization is only
produced by strongly saturated masers. A line widths analysis
indicates that the circumstellar H$_2$O masers are not saturated. Even
for large angles between the line of sight along the maser and the
direction of the magnetic field we do not expect any linear
polarization. In the LTE analysis, the lack of linear polarization can
only be explained by having the maser line of sight beam along the
magnetic field lines.

 We can compare the strength of the magnetic field in the H$_2$O maser
region with the values obtained from SiO and OH maser polarization
observations. This seems to indicate that the magnetic field strength
values inferred from the SiO maser observations are indeed due to the
normal Zeeman effect, although Elitzur (1996) has argued that the
field strength can still be a factor 10 lower on both SiO and OH
masers. Fig.2 shows the dependence of the magnetic field strength on
distance from the star. Our observed values are plotted at the
observed maximum extent of the H$_2$O maser region. Because our
observations are most sensitive for the highest magnetic fields, we
are actually probing the inner edges of the H$_2$O maser shell. The
arrows indicate the typical thickness of such a shell. These results
indicates that the magnetic field strength is best represented by a
solar-type dependence on distance ($r^{-2}$). The exact shape of the
magnetic field strength cannot easily be determined from our
observations. The SiO polarization maps indicate mostly radial field
lines close to the star.

The magnetic pressure of the field in the H$_2$O maser region
dominates the thermal pressure by a factor of 20. Using the solar-type
field, extrapolated surface field strengths are of the order of
$100-1000$ Gauss, strong enough to drive and shape the outflows.

\begin{figure}
\begin{center}
\psfig{figure=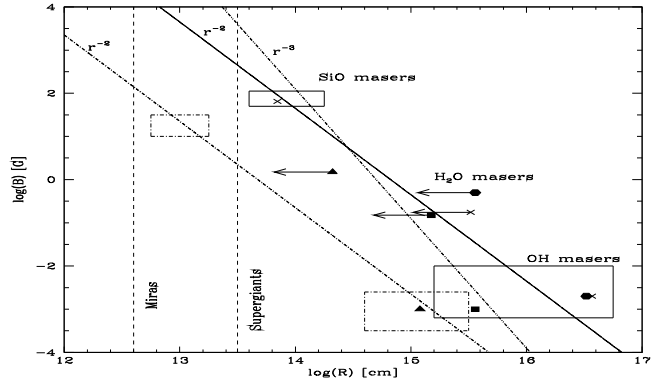,width=0.75\textwidth}
\end{center}
\caption{Magnetic field strength B, as function of distance R from the
star. Dashed-dotted boxes are the SiO and OH maser estimates for Mira
stars, solid boxes are those for supergiant stars. Symbols indicate
observations; U~Her is represented by triangles, S~Per by the square,
VY~CMa by the crosses and NML~Cyg by the hexagonal symbol. The dashed
vertical lines are an estimate of the stellar radius.}
\end{figure}


\begin{chapthebibliography}{1}
\bibitem{barv}
Barvainis, R., McIntosch, G., Predmore, C.R., 1987, Nature, 329, 613
\bibitem{black}
Blackman, E.G., Frank, A., Markiel, J.A., et al., 2001, Nature, 409, 585
\bibitem{elitz}
Elitzur, M., 1996, ApJ, 457, 415
\bibitem{nw}
Nedoluha, G.E., Watson, W.D., 1992, ApJ, 384, 185
\bibitem{sc}
Szymczak, M., Cohen, R.J., 1997, MNRAS, 288, 945
\bibitem{vlem}
Vlemmings, W., Diamond, P.J., van Langevelde, H.J., 2001, AA, 375, L1
\bibitem{vlem2}
Vlemmings, W., Diamond, P.J., van Langevelde, H.J., 2002, AA, accepted
\bibitem{ww}
Wiebe, D.S., Watson, W.D., 1998, ApJ, 503, L71

\end{chapthebibliography}
\end{document}